\documentclass[aps,prl,superscriptaddress,
amssymb]{revtex4}

\usepackage{color}
\usepackage{graphicx}
\usepackage{dcolumn}
\usepackage{bm}
\usepackage{textcomp, pifont}

\begin{document}


\title{Close proximity of FeSe to a magnetic quantum critical point as revealed by high-resolution $\mu$SR measurements}

\author{V.\ Grinenko} \email{v.grinenko@ifw-dresden.de}
\affiliation{Institute for Solid State and Materials Physics, TU Dresden, 01069 Dresden, Germany}
\affiliation{IFW Dresden, Helmholtzstrasse 20, 1069 Dresden, Germany} 
\author{R. Sarkar}
\affiliation{Institute for Solid State and Materials Physics, TU Dresden, 01069 Dresden, Germany}
\author{P.\ Materne}
\affiliation{Institute for Solid State and Materials Physics, TU Dresden, 01069 Dresden, Germany}
\author{S Kamusella} 
\affiliation{Institute for Solid State and Materials Physics, TU Dresden, 01069 Dresden, Germany}
\author{A. Yamamshita} 
\affiliation{National Institute for Materials Science (NIMS-MANA) 1-2-1 Sengen, Tsukuba 305-0047 Japan}
\author{Y. Takano}
\affiliation{National Institute for Materials Science (NIMS-MANA) 1-2-1 Sengen, Tsukuba 305-0047 Japan}
\author{Y. Sun}
\affiliation{The University of Tokyo 7-3-1 Hongo, Bunkyo-ku Tokyo 113-8656, Japan}
\author{T. Tamegai}
\affiliation{The University of Tokyo 7-3-1 Hongo, Bunkyo-ku Tokyo 113-8656, Japan}
\author{D. V.\ Efremov}
\affiliation{IFW Dresden, Helmholtzstrasse 20, 1069 Dresden, Germany}
\author{S.-L.\ Drechsler}
\affiliation{IFW Dresden, Helmholtzstrasse 20, 1069 Dresden, Germany}
\author{J.-C. Orain} 
\affiliation{Laboratory for Muon Spin Spectroscopy, PSI, CH-5232 Villigen PSI, Switzerland}
\author{T. Goko} 
\affiliation{Laboratory for Muon Spin Spectroscopy, PSI, CH-5232 Villigen PSI, Switzerland}
\author{R. Scheuermann} 
\affiliation{Laboratory for Muon Spin Spectroscopy, PSI, CH-5232 Villigen PSI, Switzerland}
\author{H. Luetkens} 
\affiliation{Laboratory for Muon Spin Spectroscopy, PSI, CH-5232 Villigen PSI, Switzerland}
\author{H.-H.Klauss} 
\affiliation{Institute for Solid State and Materials Physics, TU Dresden, 01069 Dresden, Germany}

\date{\today}

\begin{abstract}

A nematic transition preceding a long-range spin density wave antiferromagnetic phase is a common feature of many Fe based superconductors. However, in the FeSe system with a nematic transition at $T_{\rm s} \approx$ 90 K no evidence for long-range static magnetism down to very low temperature was found. The lack of magnetism is a challenge for the theoretical description of FeSe. 
Here, we investigated high-quality single crystals of FeSe using high-field (up to 9.5 Tesla) muon spin rotation ($\mu$SR) measurements. The $\mu$SR Knight shift and the bulk susceptibility linearly scale at high temperatures but deviate from this behavior around $T^{*} \sim 10$ K, where the Knight shift exhibits a kink. This behavior hints to an essential change of the electronic and/or magnetic properties crossing the region near $T^{*}$. In the temperature range $T_{\rm s} \gtrsim  T \gtrsim T^{*}$ the muon spin depolarization rate follows a critical behavior $\Lambda \propto T^{-0.4}$. The observed non-Fermi liquid behavior with a cutoff at $T^{*}$ indicates that FeSe is in the vicinity to a antiferromagnetic quantum critical point. Our analysis is suggestive for $T^{*}$ triggered by the Lifshitz transition.  
\end{abstract}


%

\maketitle 
Understanding of the interplay between nematic, magnetic and superconducting orders is one of the key problems in description of Fe based superconductors.\cite{FernandeS3014}  A lot of theoretical and experimental efforts has been applied to shed light on the nature of the broad nematic region of the FeSe system.\cite{FernandeS3014, Glasbrenner2014, Wang2015, Yamakawa2016, Xing2017} In this system superconductivity with a transition
temperature $T_{\rm c} \approx 9 K$ develops from a paramagnetic phase with a nematic transition temperature ($T_{\rm s} \approx$ 90 K) without any evidence for long-range static magnetism down to low-$T$. 

The transport properties of FeSe show a complex field and $T$ dependence below $T_{\rm s}$, \cite{Huynh2014, Watson2015a, Sun2016a, Sun2016b, Rossler2015}  which cannot only be accounted for by the anisotropic scattering in the nematic phase \cite{Tanatar2016}. The analysis of the transport data indicates essential changes of the electronic structure \cite{Huynh2014, Watson2015a, Sun2016a,  Sun2016b} and/or hints to enhanced spin fluctuations (SF) and the formation of a pseudo gap above $T_{\rm c}$. \cite{Rossler2015, Kasahara2016, Shi2018} 
Nuclear magnetic resonance (NMR) measurements revealed that FeSe exhibits a quite different SF spectrum as compared with other Fe based superconductors. The $T$ dependence of the spin-lattice relaxation rate $1/T_1T$ enhances below $T_{\rm s}$, only.\cite{Baek2015, Bohmer2015, Kasahara2016, Shi2018}  However, recent inelastic neutron scattering measurements revealed that strong N\'{e}el antiferromagnetic (AF) SF at $Q_{\rm N}$= [$\pi$, $\pi$]  above $T_{\rm s}$ exist at high energies.\cite{Wang2016a} Moreover, the SF at $Q_{\rm N}$ are suppressed and stripe AF SF at $Q_{\rm S}$= [$\pi$ ,$0$] are strongly enhanced below $T_{\rm s}$. In contrast to the  N\'{e}el SF at high temperatures the stripe SF have a noticeable low-energy tail,\cite{Wang2016b} which is accessible in the NMR experiments. In general, the experimental and theoretical investigations indicate that FeSe is on the border to magnetism. However, the energy scale which can characterize the distance to a corresponding magnetic quantum critical point (QCP) and the reasons for the pseudo gap like behavior above $T_{\rm c}$ seen in many experiments in stoichiometric FeSe are unclear. 

To get a deeper insight into the magnetic properties of FeSe in the nematic state we preformed high-resolution muon spin rotation/relaxation ($\mu$SR) experiments. The positively charged spin-1/2 muon is one of the most sensitive probes to measure local internal magnetic fields at the muon stopping sites located usually at the places of the maximal electronic density in metals. However, the exact occupation of possible stopping sites can be hardly predicted, which complicates the analysis of $\mu$SR data. The unique high-field $\mu$SR instrument in PSI Villigen allowed us to resolve the individual muon stopping sites in high-quality FeSe single crystals. This provides an access to the Knight shift and the muon depolarization rate specific to the muon site. Our local probe measurements revealed that in accord with previous results FeSe has no static magnetism in the main sample volume but low-energy SF lead to non-Fermi liquid (non-FL) behavior with a cutoff temperature $T^{*} \sim 10$ K, indicating the close proximity to a magnetic QCP. 

High-quality FeSe single crystals were grown by the vapor transport method, which is described in Refs.\cite{Sun2016a,Sun2016b} The crystals used in the $\mu$SR experiments were characterized by various techniques (see the Suppl.) For the $\mu$SR experiments the crystals with a $T_{\rm c} \approx$ 9 K were selected by magnetization measurements (Fig.\ S1 in the Suppl.). The high transversal field (TF) $\mu$SR measurements were performed at the HAL-9500 spectrometer and zero field (ZF) and low-field experiments (given in the Suppl.) were done at the DOLLY spectrometer (PSI, Villigen) on a mosaic of the selected crystals with a total mass of about 12 mg.\cite{note1} For the measurements using the HAL-9500 spectrometer the muon polarization was in the $ab$-plane. The $\mu$SR data were analyzed using the musrfit software package.\cite{Suter2012}

\begin{figure}[t]
\includegraphics[width=0.5\textwidth]{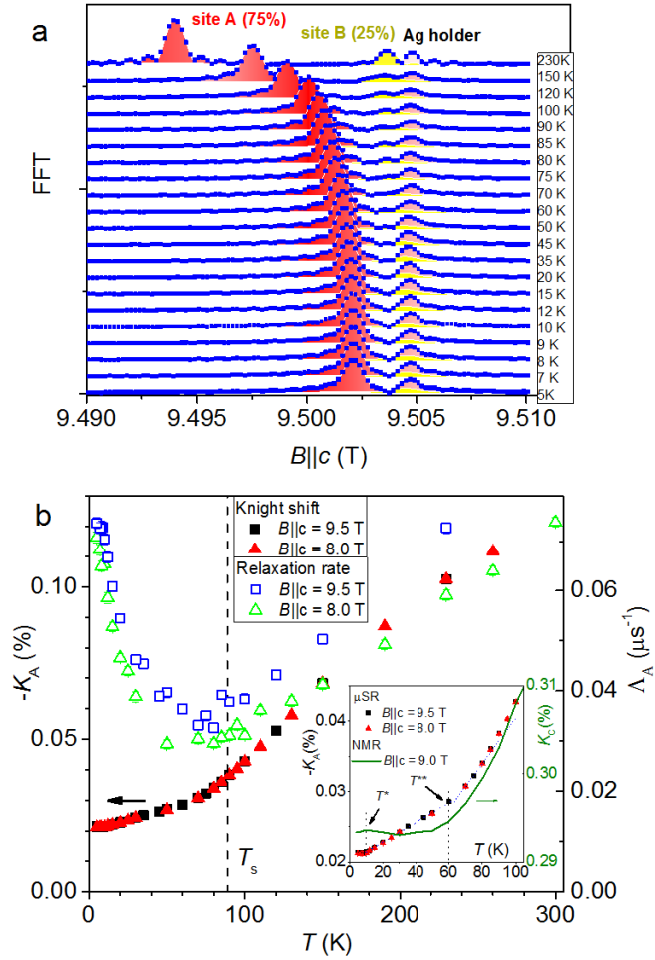}
\caption{(Color online) (a) FFT of the TF $\mu$SR time spectra measured in $B\parallel c = 9.5$ T at different temperatures for FeSe. The peaks denoted as site A and site B are related to muons stopping in the sample. The peak denoted as Ag holder is a signal from muons stopping in the silver sample holder. (b) The $\mu$SR Knight shift (left axis) and relaxation rate (right axis) of the muons stopping at site A vs. $T$ in $B\parallel c$ = 9.5 T and 8 T. Inset: the same data (left axis) and NMR Knight shift taken from Ref.\cite{Baek2015} (right axis) below 100 K.} \label{Fig:1}
\end{figure}
 
The the fast Fourier transform (FFT) of the TF-$\mu$SR spectra measured in magnetic field $B \parallel c$ = 9.5 T and at different $T$ are shown in Fig.\ \ref{Fig:1}a. The FFT spectra consist of three peaks. The right narrow peak corresponds to the Ag sample holder. Its position was used as an internal reference to calculate the absolute value of the Knight shift. The left peak (site A) corresponds to the majority of muons ($\sim 75\%$) stopping in the sample. The peak is relatively narrow, just slightly broader than the peak of the Ag but its position is strongly $T$-dependent. The position of the broad central peak (site B) is weakly $T$-dependent. By contrast, its width shows a non-monotonic $T$ dependence. The comparison with literature data \cite{Morosov1990} suggests that the muon depolarization rate for site B may be affected by the interaction of muons with the sample regions containing defects (see Figs.\ S4 and\ S5 in the Suppl.). The muons related to site B diffuse in the sample and presumably stop around diluted magnetic defects and/or magnetic sample regions. This results in a static muon depolarization rate $\Lambda_{\rm B} \sim 0.4 \mu$s$^{-1}$ of the muon spins at low-$T$ as evidenced from the comparison of the ZF and the longitudinal field $\mu$SR asymmetry spectra (Fig.\ S6 in the Suppl.). This depolarization rate corresponds to an average internal magnetic field of about 0.5 mT. According to the estimate in Ref.\ \cite{Holenstein2016} this static field might be caused by small magnetic moments in the order of $10^{-2} - 10^{-3}$ $\mu_B$ or by very diluted magnetic impurities \cite{Kirschner2016}. These estimates are also consistent with the M\"ossbauer data given in the Suppl. Figs.\ S2 and S3. This analysis imposes an upper limit for the magnetic moment of 0.03$\mu_B$ and impurity Fe concentration of 0.1 atomic $\%$. Based on both the $\mu$SR and M\"ossbauer data we exclude the presence of internal magnetic fields which could be associated with an AFM state in the  main sample volume (site A) down to $T_{\rm c}$. For further analysis we consider the signal from muons stopping at site A as representative for the main sample volume.

To obtain the Knight shift values and the relaxation rates specific to muon sites, we analyzed the high-TF $\mu$SR spectra by a sum of three contributions:
\begin{equation}
P(t) = \sum\limits_{i=1}^3 P_{\rm i}\cos\left(2\pi\nu_{\rm i} t+\frac{\pi\varphi}{180}\right){\rm exp}[-\Lambda_{\rm i}t],\label{Eq1}
\end{equation} 
where $\rm i$ corresponds to site A, site B, and Ag sample holder, respectively, and $\phi$ is the phase. The ratio $P_{\rm A}/P_{\rm B} = 3$ between the fractions of muons stopping at site A and B was found to be nearly field and $T$-independent and therefore it was fixed within the analysis. The fraction $P_{\rm Ag}$ was field dependent due to the field dependence of the beam spot size. $\Lambda_{\rm i}$ is the muon depolarization rate (peak width). The results of the fit by Eq.\ \ref{Eq1} are shown by the filling colors in Fig. \ref{Fig:1}a. The obtained $T$ dependence of the $\mu$SR Knight shift $K_{\rm A} = (\nu_{\rm A}-\nu_{\rm Ag})/\nu_{\rm Ag}$ and the muon depolarization rate $\Lambda_{\rm A}$ for site A in $B\parallel c$ = 9.5 T and 8 T are shown in Fig. \ref{Fig:1}b and for the site B in the Suppl. Fig. S4. All data shown in the paper are corrected by the demagnetization effects.\cite{note3} $K_{\rm A}$ (left axis) is field independent and shows a $T$ dependence similar to that observed in $^{77}$Se-NMR measurements as shown in the inset of Fig. \ref{Fig:1}b.\cite{Baek2015, Bohmer2015} It is seen that $K_{\rm A}$ is only slightly affected by the structural transition at $T_{\rm s}$, which results in a gradual change of the slope of the $T$ dependence. Below $T_{\rm s}$ two kink-like features are observed in the $T$ dependence of $K_{\rm A}$ around $T^{**} \sim 60$ K and $T^{*} \sim 10$ K. (We note that the feature at $T^{*}$ is not caused by the bulk superconducting transition since $T_{\rm c}$ is about 5 K in $B\parallel c$ = 9 T (inset in Fig.\ S1e in the Suppl.).) The NMR Knight shift ($K_{\rm c}$) shows a smooth variation across $T^{**}$ and it has similar to $K_{\rm A}$ behavior at $T^{*} \sim 10$ K.

\begin{figure}[t]
\includegraphics[width=0.5\textwidth]{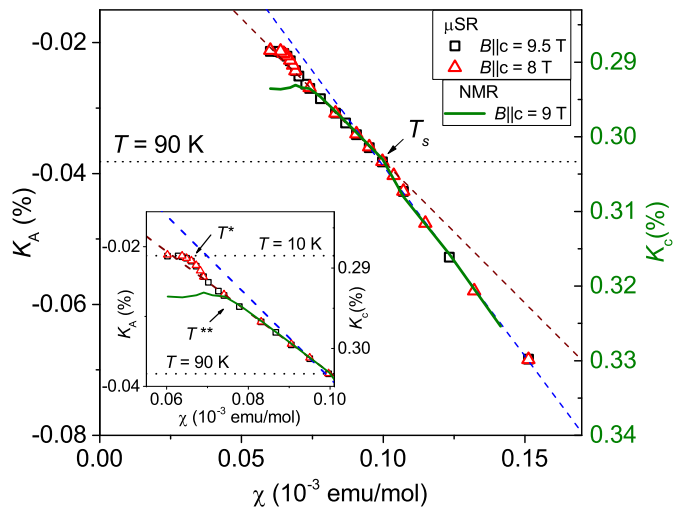}
\caption{(Color online) The $\mu$SR Knight shift $K_{\rm A}$ (Left axis) and NMR  Knight shift $K_{\rm c}$ taken from Ref.\ \cite{Baek2015} (right axis) vs. bulk molar susceptibility $\chi_{\rm m}$ of FeSe. Lines are linear fits of the data. Inset: the same data below 100 K.}
\label{Fig:2}
\end{figure}

The anomalous behavior of the electronic system at low-$T$ is seen in Fig.\ \ref{Fig:2} from the plot of $K_{\rm A}$ vs. the bulk molar susceptibility $\chi_{\rm m}$ with $T$ as an implicit parameter. The $T$ dependence of $\chi_{\rm m}$ is given in the Suppl. Figs.\ S7 - S9. In NMR experiments the Knight shift is usually linear with the susceptibility and one can extract the orbital $T$-independent contribution to the shift from the intercept and the hyperfine coupling from the slope of the linear fit.\cite{Nisson2016} At high-$T$ above $T_{\rm s}$, $K_{\rm A} = A_{\rm 0}\chi_{\rm m}+ K_{\rm orb}$ is linear with $\chi_{\rm m}$ resulting in a $T$-independent coupling constant $A_{\rm 0} \approx -3.3$ T/f.u.-$\mu_{\rm B}$ and a positive $K_{\rm orb} \approx 0.02\ \%$, which can be interpreted as a $T$-independent nuclear and orbital contribution in analogy with NMR. The parameter $A_{\rm 0}$ accounts for all $T$-dependent terms. In the case of FeSe the dipolar, hyperfine and Fermi contact contributions to the Knight shift can be $T$-dependent even above $T_{\rm s}$ due to a strong $T$ dependence of the chemical potential.\cite{Kushnirenko2017} The nematic transition results in the reduction of $A_{\rm 0} \approx -2.4$ T/f.u.-$\mu_{\rm B}$ and $K_{\rm orb}\approx 0.005\ \%$. However, $K_{\rm A}$ remains linear with $\chi_{\rm m}$. The linear relationship is violated only around $T^*$ indicating a significant $T$ variation of $A_{\rm 0}$ and $K_{\rm orb}$. The NMR $K_{\rm c}$ has the same behavior as the $\mu$SR $K_{\rm A}$ down to $T^{**}$ and becomes nearly constant at low-$T$.

The observation of kinks in the $T$ dependence of $K_{\rm A}$ (Fig.\ \ref{Fig:1}b), the broken linear relationship between $K_{\rm A}$ and $\chi_{\rm m}$ (Fig.\ \ref{Fig:2}), and features in the electrical resistivity (Fig.\ S10 in the Suppl.) and Hall coefficient \cite{Rossler2015} indicates that $T^*$ and $T^{**}$ may corresponds to a reduction of the density of states (DOS) as suggested previously in Refs. \onlinecite{Rossler2015,Kasahara2016,Shi2018}. The effects were attributed to the formation of the pseudo gap due to the charge density wave phase competing with the magnetism or preformed Cooper pairs, respectively. Alternatively, the reduction of the DOS can be caused by a $T$-induced Lifshitz transition. Indeed, a strong $T$ dependence of the Fermi energies of individual bands, was found in recent angle-resolved photoemission spectroscopy (ARPES) experiments. At high $T$ above $T_{\rm s}$ the Fermi surface (FS) consists of two hole ($h$) and two electron ($el$) like FS sheets. However, one $h$ FS pocket (around the $Z$ point) sinks below the Fermi level with the reduction of $T$ below $T_{\rm s}$ according to both ARPES and Shubnikov-de Haas oscillations measurements.\cite{Terashima2014, Watson2015b}. The situation concerning the $el$ FS sheets at low-$T$ is controversial. The Shubnikov-de Haas oscillations revealed one $el$ thin cylinder, only. However, some ARPES data indicate a complex structure around the M point of the Brillouin zone with two $el$ FS pockets. Below $T_{\rm s}$ the energy distribution curves (EDC) results in four peaks for the twinned crystals.\cite{Fedorov2016, Kushnirenko2017} In particular, one of the peak in the EDC touches the Fermi level within experimental resolution at low-$T$.\cite{Kushnirenko2017}. Therefore, the experimental data indicate a possibility of two distinct $T$-induced Lifshitz transitions below $T_{\rm s}$. 

To estimate an impact of these Lifshitz transitions on the DOS we analyzed the calculated band structure obtained using the Full Potential Local Orbital band structure package (FPLO, http://www.fplo.de) \cite{Koepernik1990}. In our calculations we observed that the experimental asymmetric shift of the Fermi energy of the $h$ and $el$ bands with $T$ results in noticeable anomalies in the DOS when the Fermi energy touches the bottom or top of the bands  (Figs.\ S11 and\ S12 in the Suppl.). The change of the DOS affects both spin and orbital contributions to the Knight shift and therefore may be responsible for the observed kinks in $K_{\rm A}$ at $T^*$ and $T^{**}$, and for the non-linear $K_{\rm A}$ vs. $\chi_{\rm m}$ plot around $T^*$ (Fig.\ \ref{Fig:2}).

\begin{figure}[t]
\includegraphics[width=0.5\textwidth]{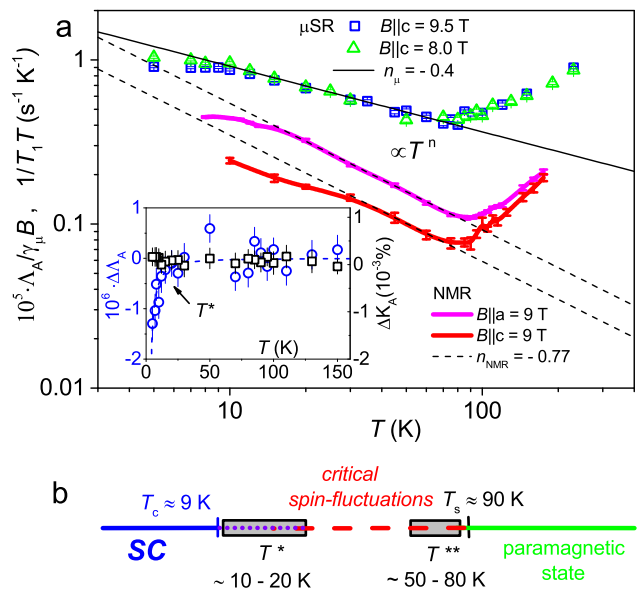}
\caption{(Color online) a) Log-log plot of the muon depolarization rate $\Lambda_{\rm A}/\gamma_{\rm \mu}B$ and the NMR $1/T_1T$ taken from Ref. \cite{Bohmer2015} vs. $T$ for FeSe. Inset: the muon depolarization rate $\Delta\Lambda_{\rm A} = \Lambda_{\rm A}(B_2)/\gamma_{\rm \mu}B_2 - \Lambda_{\rm A}(B_1)/\gamma_{\rm \mu}B_1$ (left axis), where $B_1$ = 8 T and $B_2$ = 9.5 T and the Knight shift $\Delta K_{\rm A} = K_{\rm A}(9.5{\rm T})-K_{\rm A}(8{\rm T})$ (right axis) vs. $T$. b) A schematic phase diagram of FeSe. The gray boxes show the temperature regions attributed to putative Lifshitz transitions.}
\label{Fig:3}
\end{figure}

To get a deeper insight to the effects around $T^*$ we analyzed the $T$ dependence of the muon depolarization rate $\Lambda_{\rm A}$ (Fig.\ \ref{Fig:1}b, right axis) obtained using Eq.\ \ref{Eq1}. The log-log plot of the $T$ dependence of the dimensionless $\Lambda_{\rm A}/\gamma_{\rm \mu}B$ with $\gamma_{\rm \mu} = 2\pi\times$135.53 MHz/T and that of the NMR $1/T_1T$  \cite{Bohmer2015} are shown in Fig.\ \ref{Fig:3}a. Both quantities decrease at high-$T$ and increase below $T_{\rm s}$. This behavior was interpreted as a competition between two types of SF.\cite{Hirschfeld2015} Both $\Lambda_{\rm A}$ and $1/T_1T$ for $B \parallel a$ follow a critical power-law behavior below $T_{\rm s}$ with the exponents $n_{\rm \mu} \approx -0.4$ and $n_{\rm NMR} \approx -0.77$, correspondingly. The $n_{\rm \mu} \approx -0.4$ is close to the theoretical value $-\nu(z-1)\approx -0.5$ \cite{Yaouanc2011} expected for a quantum critical behavior of itinerant AF with the correlation-length exponent $\nu = 1/2$ and the dynamical exponent $z = 2$ \cite{Leohneysen2007}. The exponent $n_{\rm NMR} \approx -0.77$ in $1/T_1T$ is close to the theoretical value $-3/4$ for 3D SDW phase found in heavy fermion systems.\cite{Kambe2009} The power-law scaling of $\Lambda_{\rm A}$ indicates that the line width in our TF data (Fig.\ \ref{Fig:1}a) is dominated by SF. This is expected in the case of a strong motional lines narrowing due to the fast diffusion of the muons in the sample, thus only muons captured around defects are affected by static line broadening (site B). In this case $\Lambda_{\rm A}(T) \propto T\int A(q)\chi(q)F(q,\omega)d^Dq$, where $A(q)$ is the $q$-dependent hyperfine coupling constant, $D$ is the dimensionality, $\chi(q)$ is the $q$-dependent susceptibility, $F(q,\omega)$ is the spectral-weight function and $\omega$ is the frequency, in our case $\omega = \gamma_{\rm \mu}B$.\cite{Yaouanc2011} Thus, it is expected that $\Lambda_{\rm A}/T \propto 1/T_1T$. The observed different exponents of $\Lambda_{\rm A}/T$ and $1/T_1T$ for $B \parallel a$, and the deviation from the power law behavior for $B \parallel c$ can be attributed to the $T$-dependent $A(q)$. For example, a strong $T$ variation of the NMR $A(q=0)$ below $T_{\rm s}$ is suggested from $K_{\rm c}$ vs. $\chi_{\rm m}$ plot (inset of Fig.\ \ref{Fig:2}).

The non-FL behavior below $T_{\rm s}$ is also evidenced by the quasi-linear dependence of the electrical resistivity $\rho$ = $\rho_0+AT+BT^2$ with $A>>BT$ (Fig.\ S10 in the Suppl.). A linear $T$-dependence is usually expected in the quantum critical region, which has a $V$-shape form in the phase diagram separated by the crossover temperature from the FL state.\cite{Shibauchi2014}.  In this sense $T^{*}$ can be interpreted as the crossover temperature. One might relate the anticipated QCP to the stripe AF order observed in FeSe under hydrostatic pressure.\cite{Wang2016, Lebert2017, Khasanov2017} Indeed, the resistivity is perfectly linear below $T_{\rm s}$ at small applied pressure $P_{\rm c} \sim 0.2 - 0.5$ GPa (see Fig.\ S13 in the Suppl. for illustration).\cite{Sun2016} Therefore, our detection of critical SF in FeSe at zero pressure suggests that the linear resistivity is related to a QCP at the pressure $P_{\rm c}$.\cite{note2} However, in the microscopic experiments such as $\mu$SR and NMR a magnetic phase under pressure is observed only at $P \gtrsim 0.8$ GPa with a relatively high transition temperature $T_{\rm N}$.\cite{Bendele2010, Wang2016, Wiecki2017} This behavior suggests that the  QCP is avoided. To settle this puzzle we propose that the QCP at zero $T$ is avoided by the same mechanism, which is responsible for the anomaly at $T^*$. 

The difference $\Delta\Lambda_{\rm A} = \Lambda_{\rm A}(B_2)/\gamma{\rm_\mu}B_2 - \Lambda_{\rm A}(B_1)/\gamma{\rm_\mu}B_1$ is shown in the inset of Fig.\ \ref{Fig:3}a, where $B_1$ = 8 T and $B_2$ = 9.5 T. $\Delta\Lambda_{\rm A}$ is almost zero above $T^*$ indicating that $\Lambda_{\rm A} \propto B$. This suggests that $F(q,\omega) \propto \omega$ at the measured frequency range, which should be related to the SF spectra of FeSe.  $\Delta\Lambda_{\rm A}$ changes the behavior across $T^*$, where the linear relationship between $K_{\rm A}$ and $\chi_{\rm m}$ is broken. The negative value of $\Delta\Lambda_{\rm A}$ indicates a relative narrowing of the lines with the magnetic field. This non-linear behavior might be caused by the changes in the shape of $F(q,\omega)$. Alternatively, it was proposed that the field dependence originates from the suppression of the $T^*$ anomaly by the magnetic field.\cite{Kasahara2016, Shi2018} The field dependence of $T^*$ is expected for both the preformed Cooper pairs and Lifshitz transition scenario (due to Zeeman effects). The formation of the Cooper pairs opens a gap at the Fermi level and therefor affects both $1/T_1T$ and Knight shift. However, $K_{\rm A}$ around $T^*$ is field independent within the error-bars (inset of Fig.\ \ref{Fig:3}a). Moreover, the anomaly in the electrical resistivity is shifted to higher $T$ with the magnetic field (Fig.\ S10 in the Suppl.). In contrast, within the Lifshitz transition scenario, the magnetic field should smear the transition and hence mainly affects the SF spectra related to the shape of the FS, whereas the DOS is less affected. A schematic phase diagram is shown in Fig.\ \ref{Fig:3}b.

We conclude that the unique high-field (up to 9.5 T) $\mu$SR investigations of high-quality single crystals of FeSe reported here allowed us to measure the local static spin susceptibility  and spin dynamics specific to particular muon stopping sites. Our microscopic investigation indicates that FeSe is close to the QCP of the SDW phase presumably located at a small pressure $P_{\rm c} \sim 0.2 - 0.5$ GPa. In contrast to usual quantum critical systems, the crossover between the quantum critical region and the low-$T$ FL state is accompanied by the reduction of the DOS. This reduction can be caused by different mechanisms such as a pseudo gap formation or the here proposed $T$-induced Lifshitz transition of the $el$ FS of the $xz/yz$ derived bands.    
 
\begin{acknowledgments}         
                          
This work was partially performed at Swiss Muon Source (S$\mu$S), PSI, Villigen. We acknowledge stimulating interest and fruitful discussion with A.\ Amato, A.\ B\"ohmer, S.\ Borisenko, M.\ Kiselev, M.\ Khodas, S.\ Kasahara, R.\ Khasanov, A. \ Maeda, S.\ Molatta, U.\ R\"ossler and Q.\ Si.  This work was supported by the DFG through grant GR 4667 and within the research training group GRK 1621. R.S. and H.H.K. are thankful to DFG for the financial assistance through the SFB 1143 for the project C02. S.-L.D. and D.E. thank the VW-Stiftung for partial support. 
\end{acknowledgments}

\section{Supplementary material}

\renewcommand{\theequation}{S\arabic{equation}}
\renewcommand{\thefigure}{S\arabic{figure}}
\renewcommand{\thetable}{S\arabic{table}}
\setcounter{equation}{0}
\setcounter{figure}{0}
\setcounter{table}{0}

In this supplementary material we provide the specific heat, M\"ossbauer, magnetization, electrical resistivity, and additional transverse field (TF), zero field (ZF), and longitudinal field (LF) $\mu$SR data of FeSe single crystals. We also present the result of our fully relativistic band-structure calculations and discuss the results in connection with available experimental angle-resolved photoemission spectroscopy (ARPES) data. The specific heat and electric transport measurements of the single crystals used in the $\mu$SR experiments were performed in a Quantum Design physical property measurement system (PPMS). The magnetization measurements were performed using a commercial superconducting quantum interference device magnetometer (SQUID) from Quantum Design.

\subsection{Basic sample characterization of FeSe single crystals}

\begin{figure}[b]
\includegraphics[width=26pc,clip]{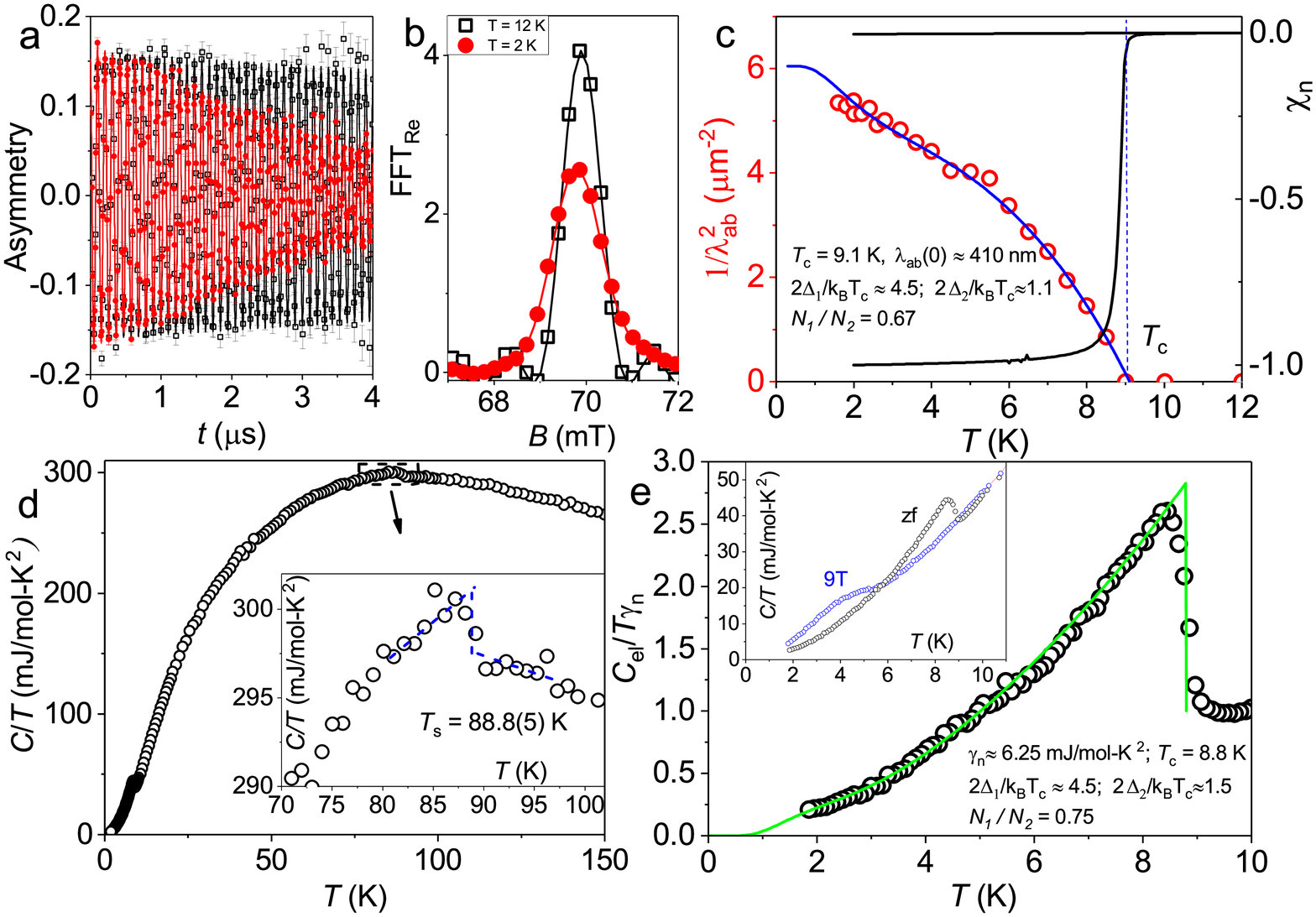}
\caption{(Color online) a) The low transverse-field $\mu$SR time spectra of FeSe  above and below $T_{\rm c}$ measured in applied field $B\parallel c = 70$ mT. b) Real part of the FFT of the data shown in Fig.\ref{Fig:1}a. c) Temperature dependence of the penetration depth (left axis) and normalized susceptibility (right axis) measured in $B\parallel c = 0.5$ mT of the same samples. d) Temperature dependence of the specific heat of the FeSe single crystal measured in zero magnetic field. Inset: zoom of the data around the nematic transition. e) Temperature dependence of the electronic specific heat at low temperatures. Inset: low temperature part of the total specific heat measured in zero magnetic field and $B\parallel c = 9$ T. } 
\label{FigS:1}
\end{figure}

The $\mu$SR time spectra of the FeSe single crystals above and below $T_{\rm c}$ measured in a transverse magnetic field (TF-$\mu$SR) $B \parallel c$ = 70mT applied parallel to the crystallographic $c$-axis are shown in Fig.\ \ref{FigS:1}a. The real part of the fast Fourier transform (FFT) of the spectra is shown in Fig.\ \ref{FigS:1}b. The normal state data form a narrow line  indicating the absence of any noticeable magnetic impurities or static magnetism. A high crystalline quality and very low amount of impurities in our FeSe sample are also evidenced by narrow M\"ossbauer spectra with Fe impurity below 0.1 atomic $\%$ (Fig.\ \ref{FigS:2}). The strong damping in the superconducting state indicates the formation of a vortex lattice in the whole sample volume in accord with the bulk superconductivity demonstrated by the specific heat data (Fig.\ \ref{FigS:1}e). The temperature dependence of the  penetration depth $\lambda_{\rm ab}$ obtained from the second moment of the field distribution \cite{Maisuradze2009} is shown in Fig.\ \ref{FigS:1}c (left axis). The $T_{\rm c}$ value estimated from the $\mu$SR measurements is in a good agreement with the values from the magnetization measurements (right axis in Fig.\ \ref{FigS:1}c) and the specific heat.(Fig.\ \ref{FigS:1}e) The fitting of the penetration depth and the specific heat data using a minimal two-band $\alpha$-model with two BCS-like isotropic $s$-wave gaps $\Delta_{\rm i}$, having different DOS $N_{\rm i}$, results in two distinctly different values of the gaps as shown in Fig.\ \ref{FigS:1}. These values are in a good agreement with previous $\mu$SR data obtained on a polycrystalline sample with slightly lower $T_{\rm c}$.\cite{Khasanov2008} We note that the real gap structure is more complex than that implemented in our $\alpha$-model.\cite{Sun2017} However, a more sophisticated analysis requires experimental data down to mK temperatures which is beyond the scope of the present study. Our crystals also show a clear anomaly in the specific heat at $T_{\rm s}$ (Fig.\ \ref{FigS:1}d) in good agreement with previous data.\cite{Rossler2015}

\subsection{M\"ossbauer spectra of FeSe single crystals}

\begin{figure}[t]
\includegraphics[width=20pc,clip]{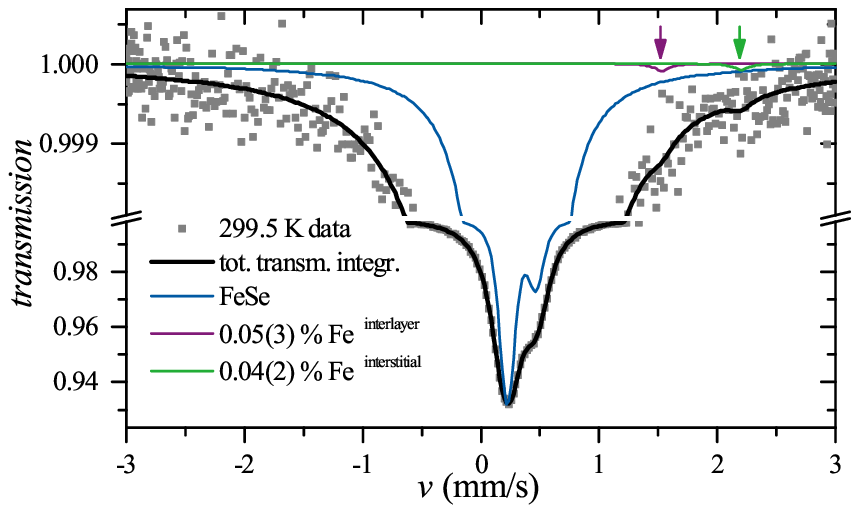}
\caption{(Color online) High statistics M\"ossbauer spectrum proves the outstanding sample quality. Typical secondary iron lines do not contribute significantly. The data indicate a very small amount of impurity Fe. Note different scales for the transmission intensity at the central peak and the shoulders.} 
\label{FigS:2}
\end{figure}

\begin{figure}[t]
\includegraphics[width=20pc,clip]{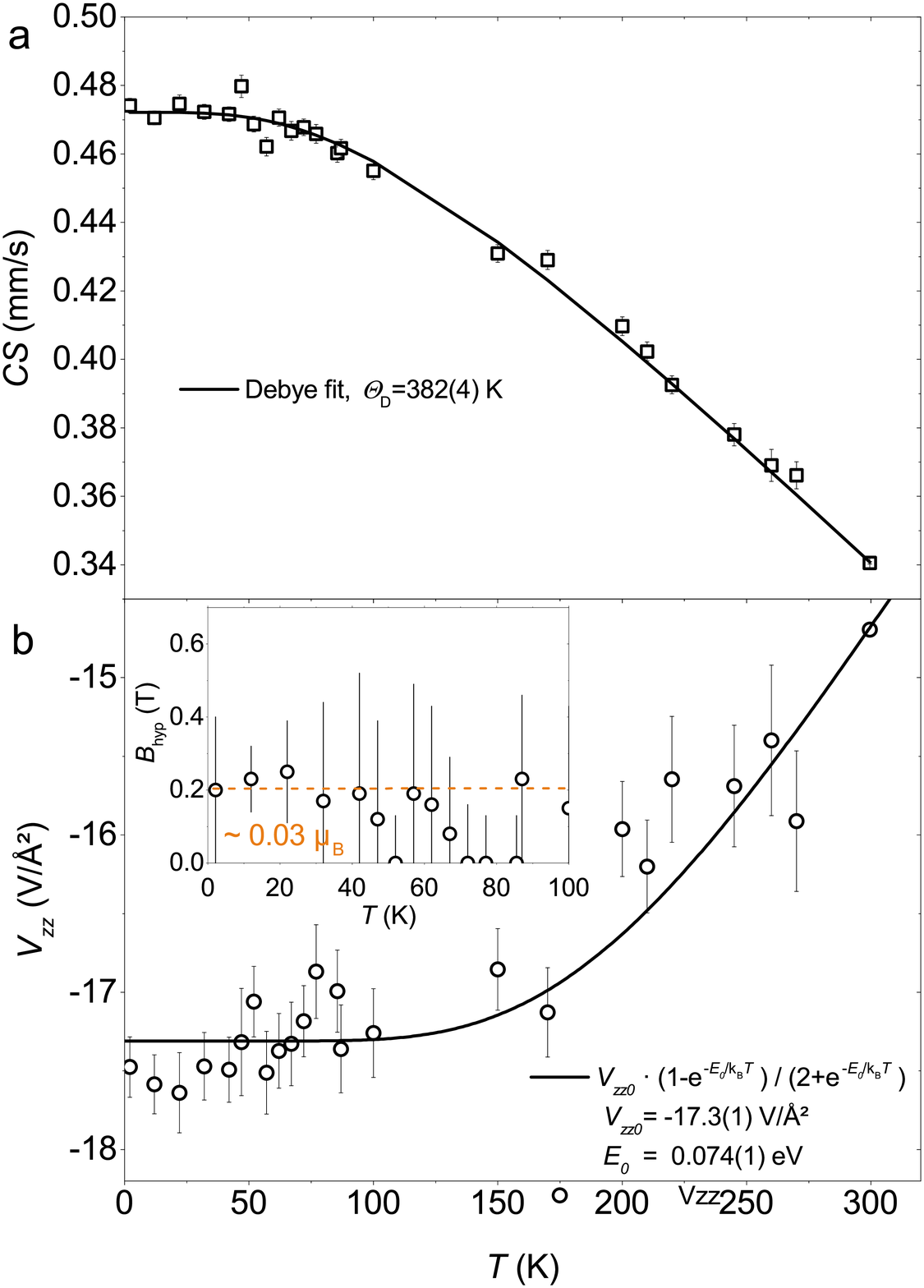}
\caption{(Color online) a) Temperature dependence of the center shift. The solid line is a theoretical curve of the quadratic Doppler effect within the Debye approximation. b) Temperature dependence of the principal component
of the electric field gradient. The solid curve is a fit for tetrahedral Fe environment according to Ref. \cite{Shylin2015}. Inset: the difference between the experimental data and the theoretical curve shown in Fig.\ \ref{Fig:3}b recalculated in Tesla assuming that the difference is related to a contribution due to the hyperfine field.} 
\label{FigS:3} 
\end{figure}

A high statistic M\"ossbauer spectrum with a baseline of $10^{7}$ counts was taken at 299.5 K (Fig.\ \ref{FigS:2}). Assuming a Debye temperature of 382 K
from the Quadratic Doppler effect (Fig.\ \ref{FigS:3}a), the Debye Waller factor at room temperature is $f_{\rm a}$ = 0.747. Consequently, $t_{\rm a}$ = 7.1 was assumed for the transmission integral with a resonant fraction $f_{\rm r}$ as a free fit parameter. However, this high $t_{\rm a}$ models the spectrum of an 
thick absorber, which both concerning line leveling and width exceeds the measured spectrum. If both $f_{\rm r}$ and $t_{\rm a}$ were used as free fit parameters and a small canting of $\beta = 5^o$ of the crystals is assumed, $t_{\rm a}$ = 3.1 results. This corresponds to a sample mass of 5 mg (the actual sample mass is about 12 mg). In such a single spectrum fit the absorber line width $\omega_{\rm abs}$ = 0.0643(4) is close to natural line width and rarely
 observed. However, this Lorentzian broadening yields only $\chi^2_{\rm red} = 1.44$, with a full Voigt line shape it is decreased to $\chi^2_{\rm red} = 0.98$. The observed effective principal component of the electric field gradient
 (EFG) $V_{\rm zz,eff}$ = -14.57(4) V/${\rm \AA^2}$ a center shift of +0.450(1) mm/s with respect to room temperature iron is in good agreement with previous 
studies \cite{Blachowski2009,McQueen2009}. The Voigt lines shape is used to search for any additional contributions to the spectrum. In that case additional lines are expected at 2.2 mm/s \cite{Friedrichs2015} and eventually at 1.5 mm/s \cite{Pachmayr2015}. The corresponding fractions are 0.0005(3) and 0.0004(2). In terms of M\"ossbauer spectroscopy an upper limit of $0.2\%$ of secondary iron can be concluded.

No obvious changes of the spectra with temperature were observed. A small increase of the quadrupole splitting with lowering temperature (Fig.\ \ref{FigS:3}b) following Ref. \cite{Shylin2015} was attributed to a typical  behavior of the distorted tetrahedral environment of the $^{57}$Fe nucleus. The crystal field splitting of 0.074(1) eV extracted from this model is consistent with the value 0.72(2) eV obtained by Shylin {\it et al}. \cite{Shylin2015}. The difference between the experimental data and the theoretical model gives an upper limit for the possible hyperfine magnetic fields of about 0.2 T at low temperatures (see the inset of Fig.\ \ref{FigS:3}b). For typical Fe based superconductors this corresponds to 0.03 $\mu_{\rm B}$ of the quasi-static Fe magnetic moments. 

\newpage
\subsection{Additional $\mu$SR data of FeSe single crystals}

The experiments at the DOLLY spectrometer were performed in transverse and longitudinal polarization. In the transverse polarization mode the muon spin polarization is rotated by $45^o$ with respect to the direction of the beam and along the beam direction for the longitudinal polarization mode.

\begin{figure}[t]
\includegraphics[width=20pc,clip]{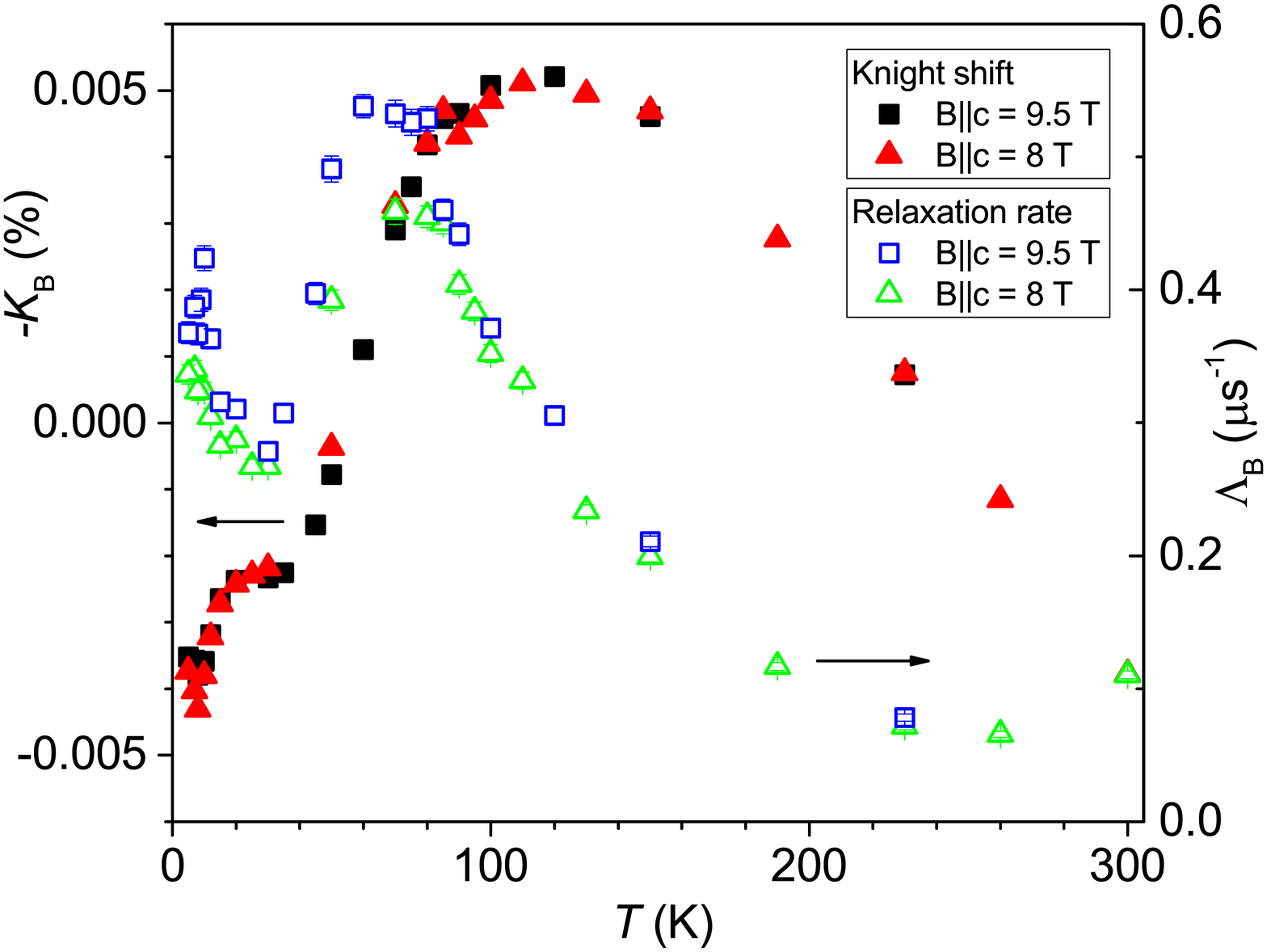}
\caption{(Color online)  Temperature dependence of the $\mu$SR Knight shift (left axis) and relaxation rate (right axis) of the muons stopping at the Site B measured in $B\parallel c$= 9.5 T and 8 T. The notations are the same as in the main text.} 
\label{Fig:4}
\end{figure}

\begin{figure}[b]
\includegraphics[width=20pc,clip]{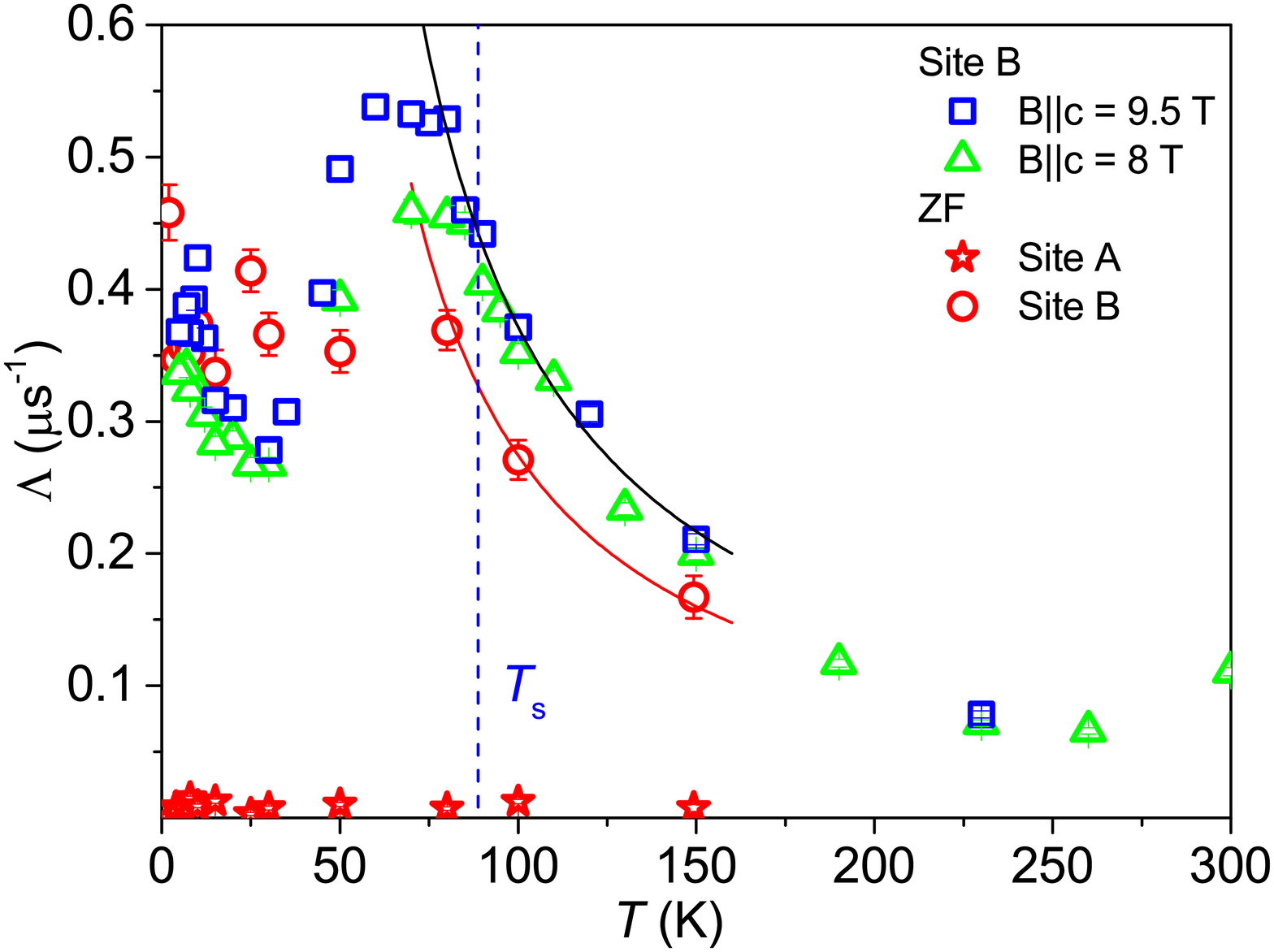}
\caption{(Color online) Temperature dependence of the $\mu$SR relaxation rate of FeSe single crystals obtained from zero field ZF and TF measurements (the same data as in Fig.\ \ref{Fig:4}).} 
\label{Fig:5}
\end{figure}

\begin{figure}[t]
\includegraphics[width=20pc,clip]{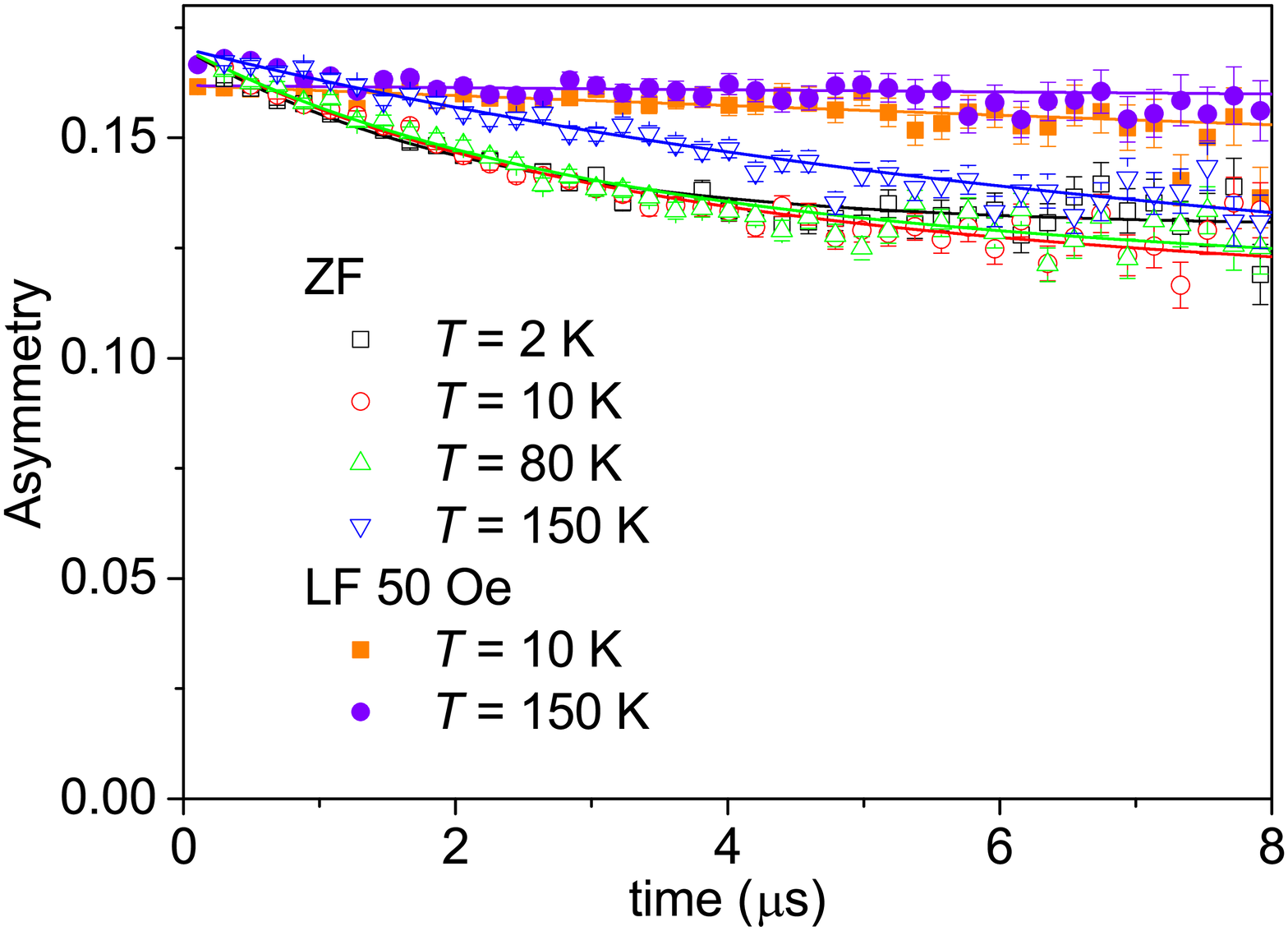}
\caption{(Color online) Representative ZF-$\mu$SR asymmetry time spectra of FeSe single crystals measured in the transverse polarization mode, and LF spectra measured in the longitudinal polarization mode (multiplied by cos$(45^o)$) to fit the initial asymmetry values in the transverse polarization mode. The solid curves are fits to Eq.\ \ref{Eq1}.} 
\label{Fig:6}
\end{figure}

To fit the ZF and LF asymmetry spectra the same model (with two muon sites) as in the case of the high-TF measurements was used:
\begin{equation}
A(t) = [1- A_{\rm bg}][A_{\rm A}(0){\rm exp}[-\lambda_{\rm A} t]+A_{\rm B}(0){\rm exp}[-\lambda_{\rm B} t]]+A_{\rm bg},\label{EqS1}
\end{equation} 
where $A_{\rm A}= 0.75$ and $A_{\rm B} = 0.25$ are the initial sample asymmetries, $\lambda_{\rm A}$ and $\lambda_{\rm B}$ are the relaxation rates, and $A_{\rm bg} = 0.16$ is the non-relaxing background asymmetry obtained from the fit of low-TF data shown in Fig.\ 1 (main text) as described in Ref. \onlinecite{Grinenko2017}. The results of the fit by Eq.\ \ref{EqS1} are shown by the solid lines in Fig. \ref{Fig:6}. 

\newpage

\subsection{Magnetization data of FeSe single crystals}

\begin{figure}[b]
\includegraphics[width=20pc,clip]{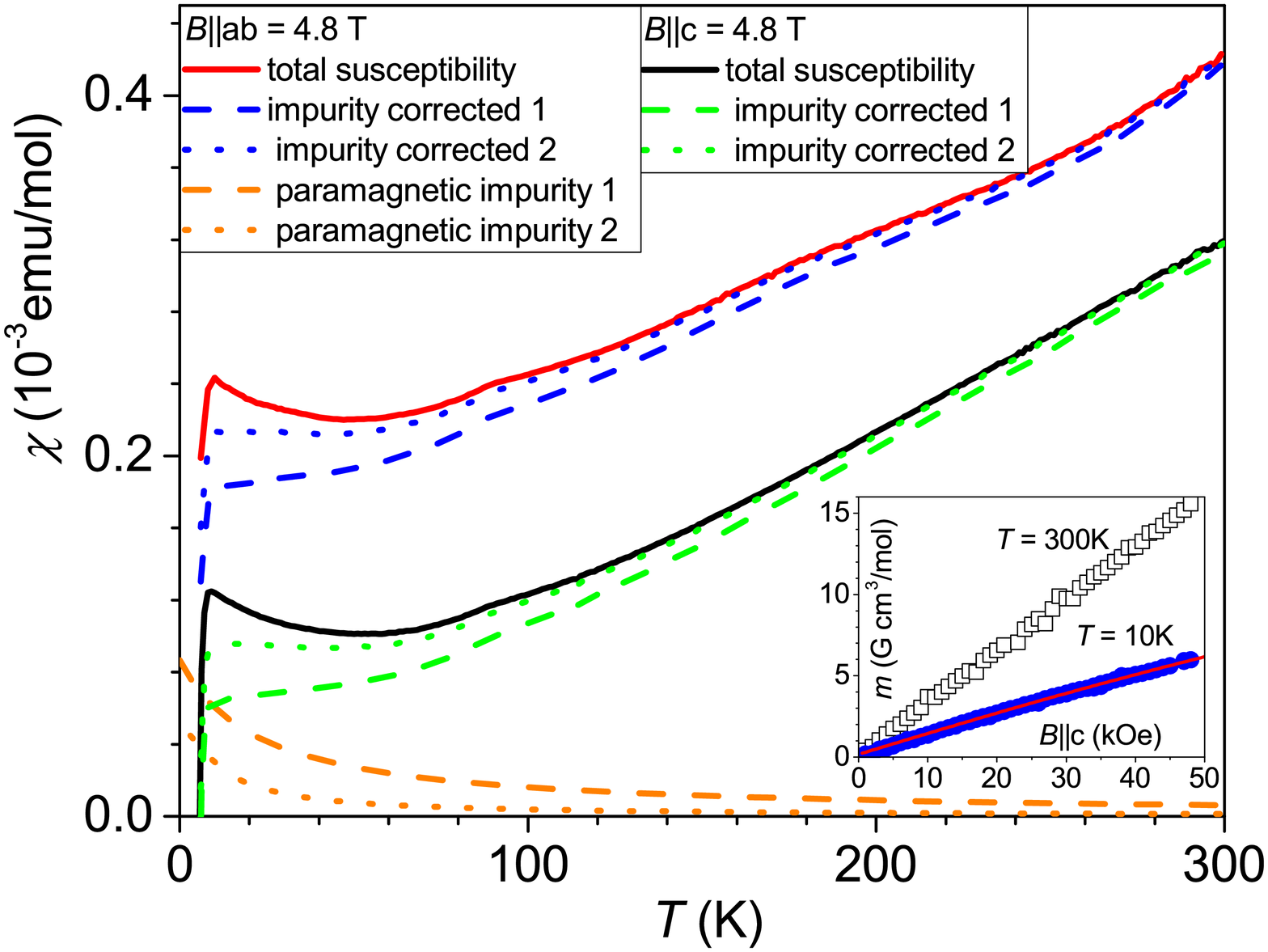}
\caption{(Color online) Temperature dependence of the molar
susceptibility $ \chi_m$ of the FeSe single crystal samples used in the $\mu$SR experiments - solid curves. The dashed and dotted curves are corrected susceptibility to paramagnetic impurity contribution using different methods: impurity correction 1 (dashed lines) is obtained as shown in Fig.\ref{Fig:8}, impurity correction 2 (doted lines) is the paramagnetic impurity contribution obtained from the fit of the magnetization data shown in the inset using the Brillouin function for the magnetization of the diluted impurities.} 
\label{Fig:7}
\end{figure}

The magnetization data given in the inset of Fig. \ref{Fig:7} were fitted using two 
contributions: $m(H,T) = m_{\rm int}(H,T)+m_{\rm imp}(H,T)$, where $m_{\rm int}$ is the intrinsic magnetization and $m_{\rm imp}$ is the magnetic impurity contribution. To estimate the concentration of the magnetic impurities $n$ in the samples we assumed that the impurity contribution is related to paramagnetic 
Fe$^{+2}$ with $J$=4.\cite{Grinenko2014} We note that the particular type of the impurity is not essential for the evaluation of the intrinsic magnetization and affects mainly the estimated amount of the impurities $n$. In this case $m_{\rm imp} = nJgB(H/T)$, where $B(H/T)$ is the Brillouin function, and $g$ is the
Land\' e  factor. From the fit of the experimental data shown in the inset of Fig. \ref{Fig:7} we estimated that $n \lesssim 0.1{\rm mol}\%$ for the investigated samples in qualitative accord with the M\"ossbauer data shown in Fig. \ref{Fig:2}.

\begin{figure}[t]
\includegraphics[width=20pc,clip]{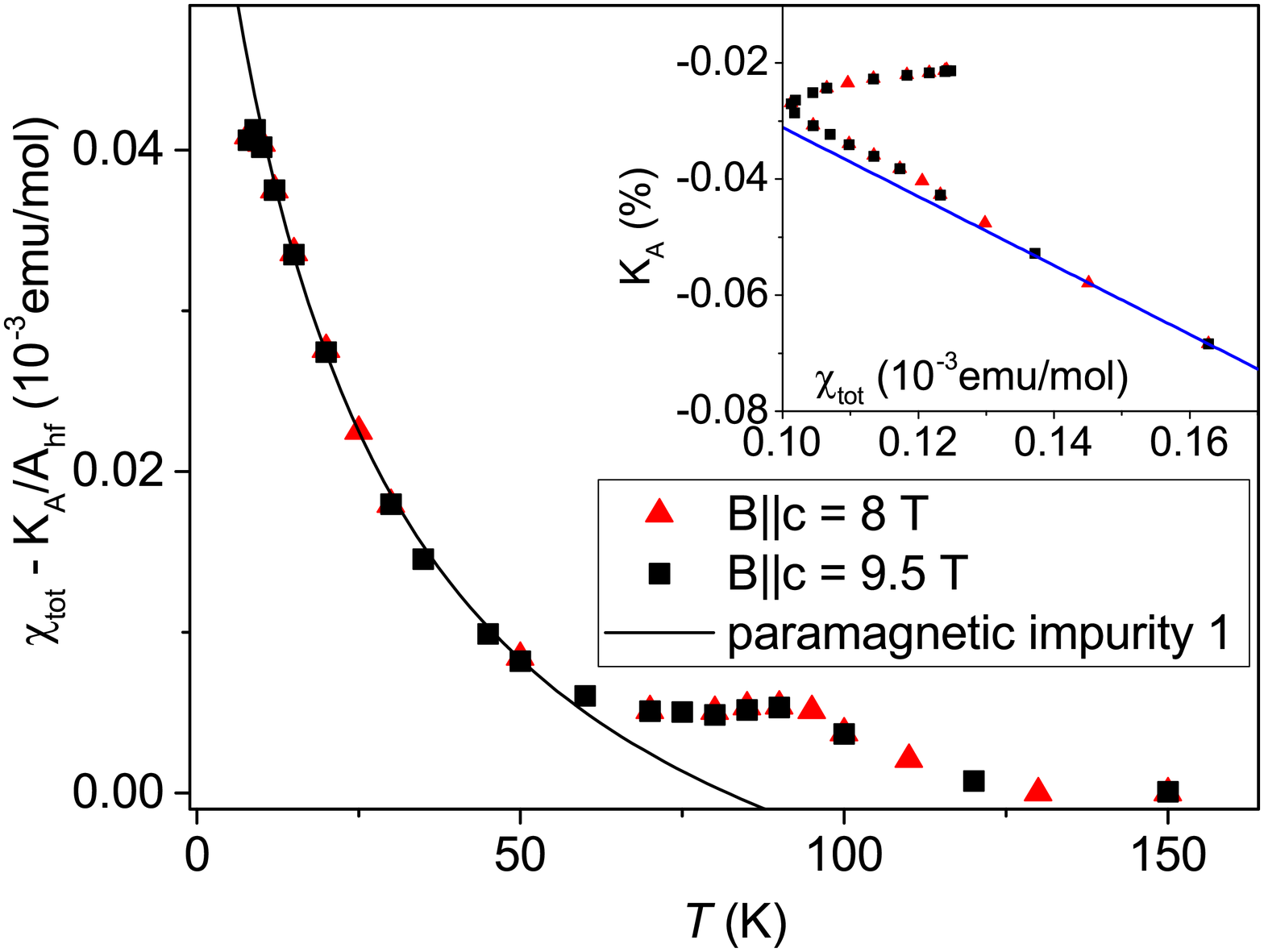}
\caption{(Color online) Temperature dependence of the difference between the bulk magnetic susceptibility $\chi_{\rm tot}$ (Fig.\ref{Fig:7}) and $K_{\rm A}/A_{\rm hf}$. The $A_{\rm hf}$ value is obtained at high temperatures from the linear fit of the $K_{\rm A}$ data plotted versus $\chi_{\rm tot}$ (inset). In this analysis is assumed that the impurity contribution to the bulk susceptibility is very small at high temperatures, whereas the difference $\chi_{\rm tot} - K_{\rm A}/A_{\rm hf}$ at low temperatures is dominated by the impurity contribution. The Curie-Weiss fit shown in main figure gives the approximate impurity contribution to the susceptibility $\chi_{\rm tot}$. This correction is used in Figs.\ref{Fig:7} and \ref{Fig:9} as the paramagnetic impurity 1.} 
\label{Fig:8}
\end{figure}

\begin{figure}[t]
\includegraphics[width=20pc,clip]{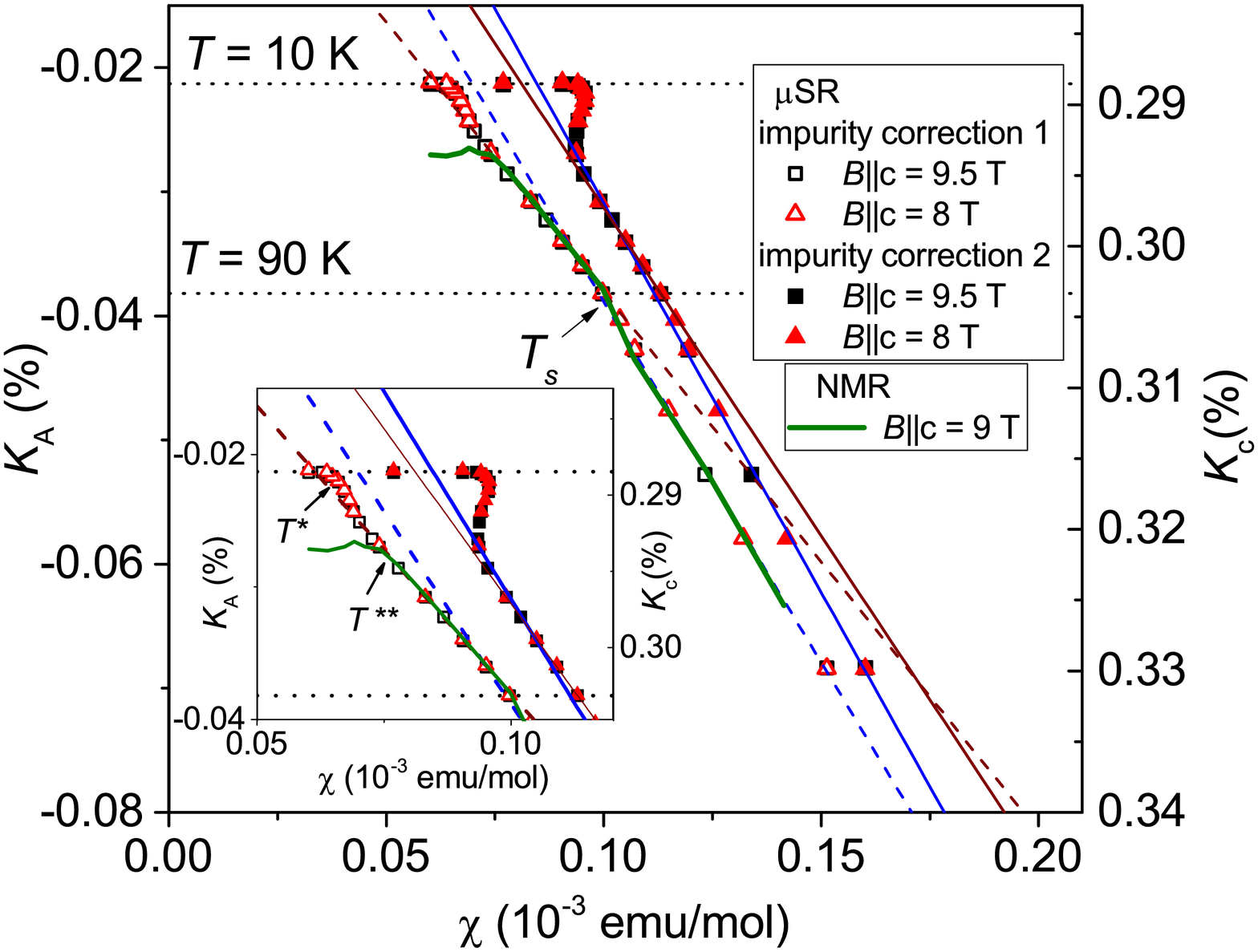}
\caption{(Color online) The $\mu$SR Knight shift $K_{\rm A}$ (Left axis) and NMR  Knight shift (right axis) vs. the  bulk molar susceptibility $\chi_{\rm m}$ of FeSe single crystals. $\chi_{\rm m}$ is corrected to the paramagnetic impurity contribution obtained as shown in Figs. \ref{Fig:7} and \ref{Fig:8}. In spite of a different corrections the temperature position of the features around $T^*$ and $T^{**}$ discussed in the main text is unchanged. The NMR data are taken from Ref.\ \cite{Baek2015}. Lines are linear fits of the data. Inset: the same data below 100 K.} 
\label{Fig:9}
\end{figure}

\newpage

\subsection{Electrical resistivity of FeSe single crystals}

\begin{figure}[t]
\includegraphics[width=30pc,clip]{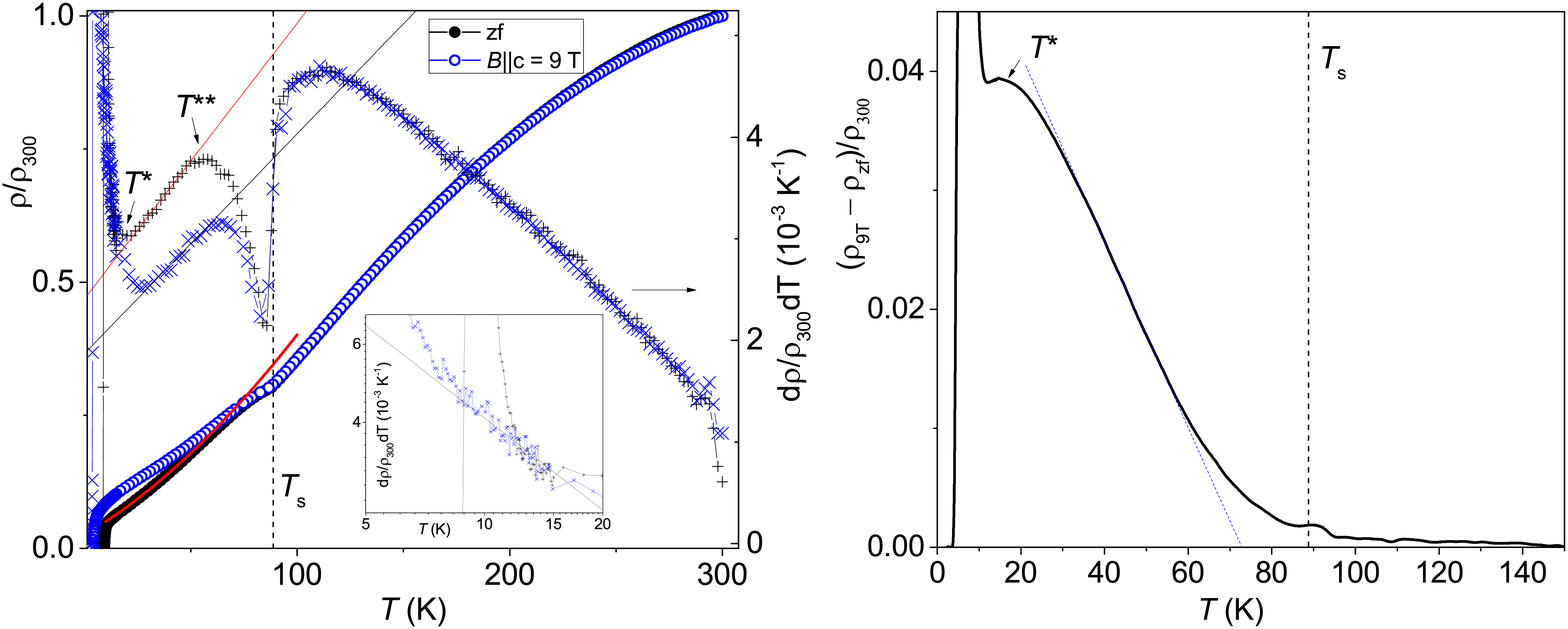}
\caption{(Color online) a) Temperature dependence of the electrical resistivity of the FeSe single crystal (left axis) and its derivative (right axis) in zero and applied magnetic field $B\parallel c = 9$ T. b) Temperature dependence of the magnetoresistance of the same crystal. The notations are the same as in the main text.} 
\label{Fig:10}
\end{figure}

Temperature dependence of the normalized electrical resistivity $\rho/\rho_{\rm 300}$ of the FeSe single crystal is shown in Fig.\ \ref{Fig:10}a (left axis) and its temperature derivative (right axis). The magneto resistance $(\rho_{\rm 9T} - \rho_{\rm ZF})/\rho_{\rm 300}$ is shown in Fig.\ \ref{Fig:10}b, where $\rho_{\rm 9T}$ is the resistance at 9T, $\rho_{\rm ZF}$ is the ZF resistance, and $\rho_{\rm 300}$ is the resistance at $T$ = 300 K. 

\newpage
\subsection{Band structure of FeSe below $T_{\rm s}$}

\begin{figure}[t]
\includegraphics[width=18pc,clip]{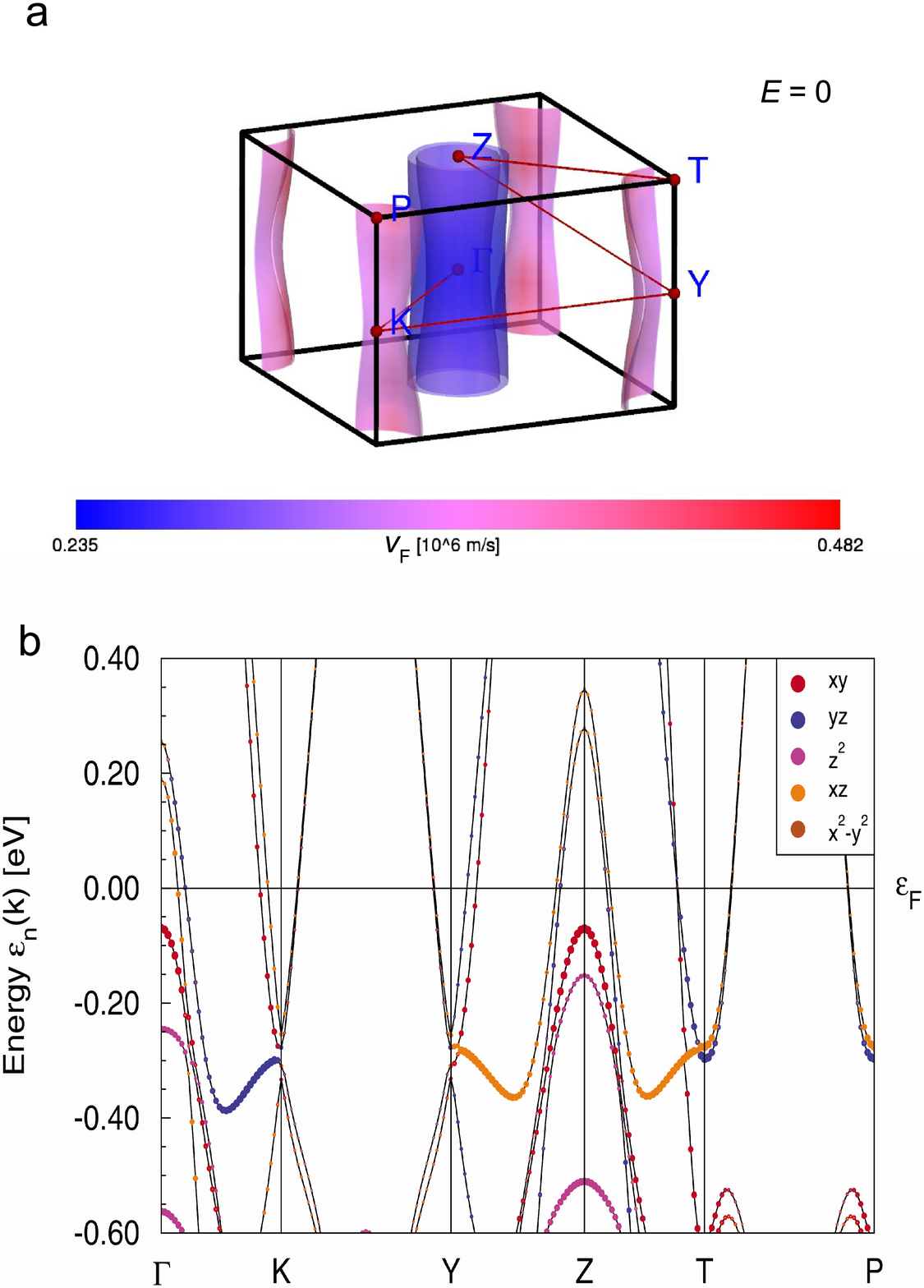}
\caption{(Color online) a) The calculated Fermi surface ($E =$ 0 is the Fermi energy) of FeSe using orthorhombic lattice parameters from Ref.\cite{Khasanov2010}. b) The calculated band structure of FeSe.} 
\label{Fig:11}
\end{figure}

\begin{figure}[t]
\includegraphics[width=20pc,clip]{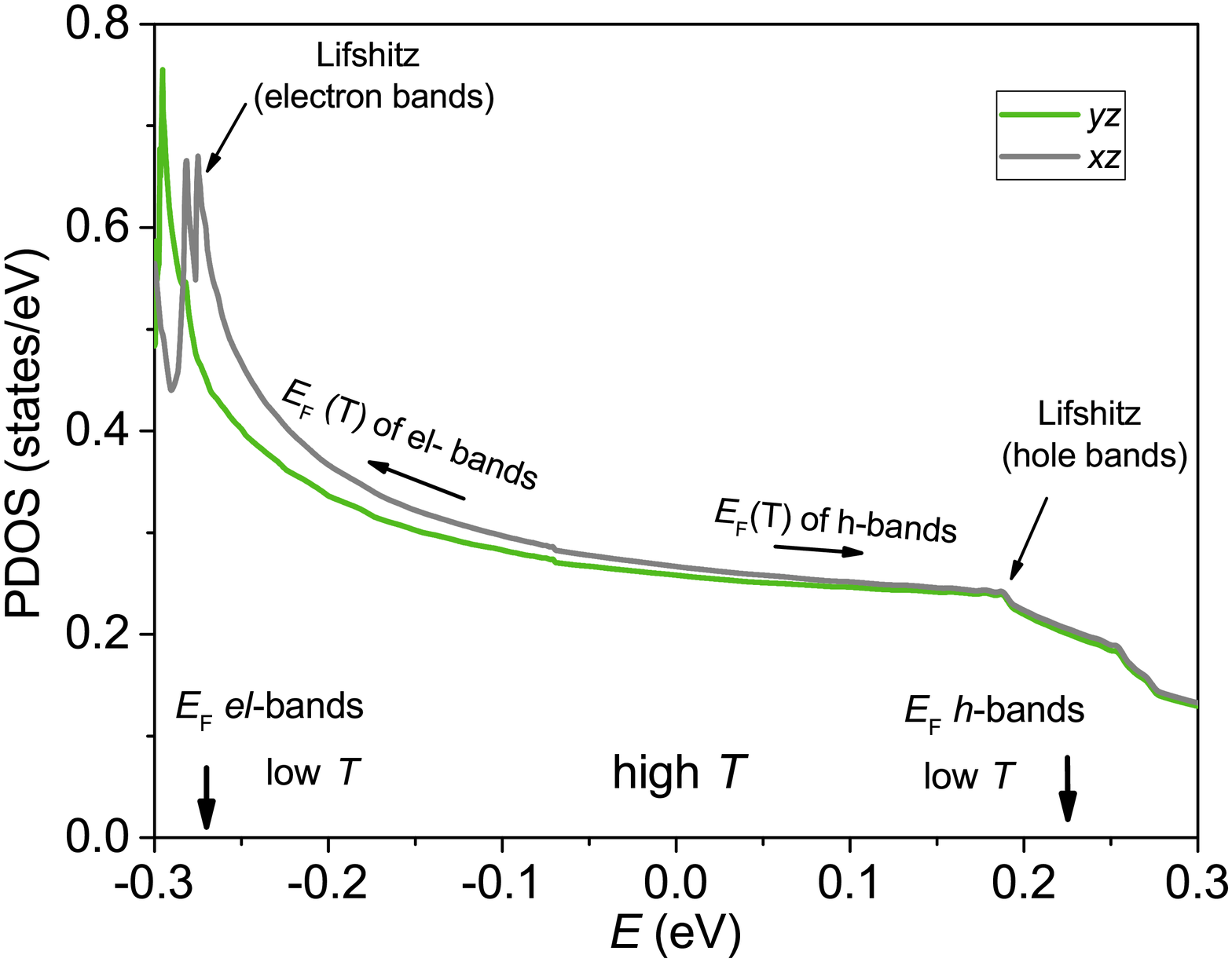}
\caption{(Color online) The partial density of states of FeSe close to the chemical potential for $xz$ and $yz$ orbitals. The position of the experimental Fermi level is different from the calculated one due to the temperature and band dependent {\it blue-red} shifts of the Fermi energy. The arrows show the direction of the Fermi level shift with temperature for electron and hole bands according to ARPES experiments\cite{Kushnirenko2017}.} 
\label{Fig:12}
\end{figure}

To sketch qualitatively (i.e. ignoring the strong $el-el$ interaction effects)  the impact of the temperature and band dependent {\it blue-red}  shifts on the Fermi surfaces \cite{Kushnirenko2017}, we performed fully relativistic density functional theory (DFT) calculations of the band structure for FeSe bulk single crystals. Our calculations were carried out within the generalized gradient approximation GGA using the Full Potential Local Orbital band structure package (FPLO, http://www.fplo.de)\cite{Koepernik1990}. A $k$-mesh of 72x72x36 k-points in the whole Brillouin zone was employed. The calculations were performed using the FPLO ab-initio Simulation Package within the Perdew, Burke and Ernzerhof (PBE) functional for the exchange-correlation potential. The calculated Fermi surface, the band structure, and the partial density of states summarized in Figs. \ref{Fig:11} and \ref{Fig:12}. According to ARPES data \cite{Kushnirenko2017, Watson2015b} the experimental position of the Fermi level is close to the theoretical predictions at high temperatures (above the structural transition temperature) but it is essentially different at low temperatures due to the antisymmetrical, so called, {\it blue-red} shift of the individual Fermi levels for electron and hole bands. Using the experimental position of the Fermi levels at low temperatures \cite{Fedorov2016} one can see that two Lifshitz transitions occur with the reduction of the temperature below $T_{\rm s}$. The first Lifshitz transition for the central hole Fermi surface occurs around the $Z$-point of the Brillouin zone and the second Lifshitz transition may occur very close to $T_{\rm c}$ involving parts of the electron Fermi surface.\cite{Kushnirenko2017}

\newpage
\subsection{Tentative phase diagram of FeSe under hydrostatic pressure.}      

\begin{figure}[t]
\includegraphics[width=20pc,clip]{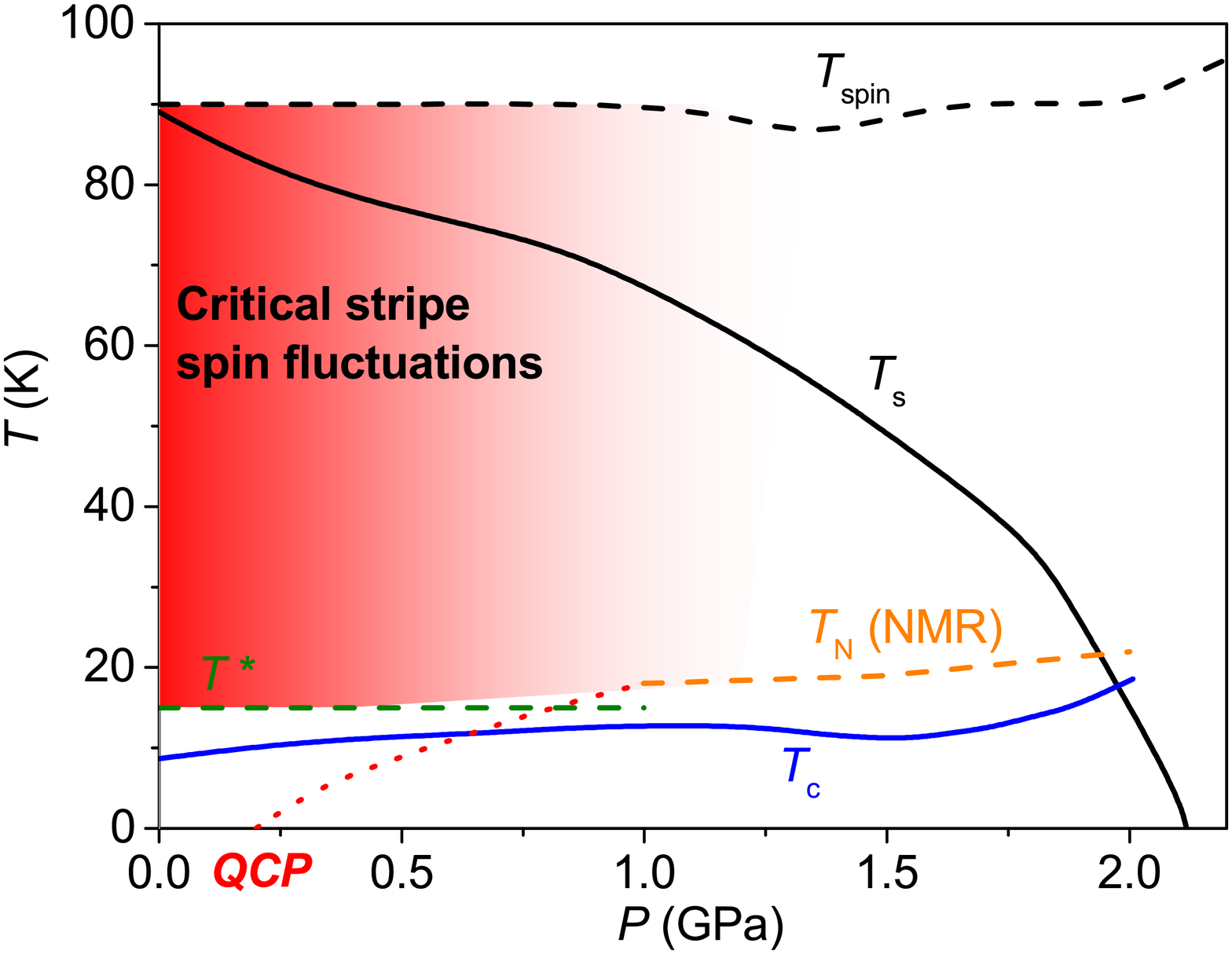}
\caption{(Color online) The tentative phase diagram of FeSe under hydrostatic pressure. Red color shows the expected quantum critical region above anomaly at $T^*$ defined in the main text. $T_{\rm spin}$ is the onset of the magnetic fluctuations measured by the NMR \cite{Wang2016}, $T_{\rm s}$ is the nematic transition temperature \cite{Sun2016}, $T_{\rm N}$ is the antiferromagnetic transition temperature \cite{Wiecki2017}, $T_{\rm c}$ is the superconducting transition temperature \cite{Sun2016}, and  the red dotted line shows the putative position of the avoided quantum critical point (QCP).} 
\label{Fig:13}
\end{figure}

Based on our analysis (see also the main text) we
predict that $T^*$ is nearly pressure independent up to $P \sim 1$ GPa. Above $P \sim 1$ GPa the N\'{e}el temperature $T_{\rm N} > T^*$. The antiferromagnetic transition results in the reconstruction of the Fermi surface, which should suppress anomaly at $T^*$. The tentative phase diagram of FeSe is shown in Fig.\ \ref{Fig:13}.

\begin{thebibliography}{100}
\bibitem{note3} The $K_{\rm dem} = 4\pi(1/3-N)\rho_{\rm m} \chi_{\rm m}$ is the correction for the demagnetization and Lorentz fields (where $N \approx 1$ is the demagnetization factor and $\rho_m = 0.0426$ mol/cm$^3$ is the molar density).
\end{thebibliography}
\end{document}